\documentclass[12pt]{article}
\usepackage{cite,amssymb,amsmath}
\usepackage{graphicx,graphics}
\begin{document}
\title{\bf Hierarchy of Fermion Masses in the 5D Standing Wave Braneworld}
\author{{\bf Merab Gogberashvili}\\
E. Andronikashvili Institute of Physics\\
6 Tamarashvili Street, Tbilisi 0177, Georgia\\
and\\
I. Javakhishvili Tbilisi State University\\
3 Chavchavadze Avenue, Tbilisi 0179, Georgia\\
{\sl E-mail: gogber@gmail.com}
\\
\\
{\bf Giorgi Tukhashvili}\\
I. Javakhishvili Tbilisi State University\\
3 Chavchavadze Avenue, Tbilisi 0179, Georgia\\
{\sl E-mail: gtukhashvili10@gmail.com}
\date{October 10, 2012}
}
\maketitle
\begin{abstract}
The localization problem for massive fermions in the 5D standing wave braneworld is considered. The fermion generations on the brane are explained by the existence of three nodes of standing waves in the bulk, where the left and right modes of the 5D fermion are localized and Dirac-type couplings mixes them. We show that in this model the structure of peaks of the left and right fermion wavefunctions in extra dimension leads to the observed mass spectrum of the up and down quark families.

\end{abstract}

\vskip 0.5cm

%%%%%%%%%%%%%%%%%%%%%%%%%%%%%%%%%%%%%%%%%%%%%%%%%%%%%%%%%%%%%%%%%%%%%%%%%%%%%%%

The Standard Model (SM) of particle physics is very successful and has been tested by experiments with high precision. However, from the theoretical perspective, there still exist several puzzles, such as the origin of the fermion masses, their mixing and three generations structure. At present it seems that we do not have satisfactory theoretical framework to explain these phenomena and an extensions of the SM is believed to be necessary. There have been several approaches to address the above problem beyond SM, a popular way is the introduction of the family symmetry \cite{Family}.

Large extra dimensions have also been proposed to address problems in flavor physics (see, \cite{Split,m-mix} and references therein). For example, in the 'Split Fermions' model \cite{Split} the SM fields are confined to a thick brane with the fermions localized in specific points in the bulk. If the fermion wavefunctions are highly peaked in the extra dimension, a small relative shift between the peaks leads to a large suppression of the effective Yukawa coupling due to the smallness of the overlap of the wavefunctions. In this paper we consider a similar model within the 5D standing wave braneworld scenario \cite{Wave,JHEP}. We connect fermion families with the existence of nodes of standing waves in the bulk where various modes of the single 5D fermion field reside.

The 5D standing wave braneworld is formed by the background solution to the coupled Einstein and massless ghost scalar field equations \cite{Wave,JHEP}:
\begin{equation} \label{metric}
ds^2 = e^{2a|r|}\left( dt^2 - e^{u}dx^2 - e^{u}dy^2 - e^{-2u}dz^2 \right) - dr^2~,
\end{equation}
where $r$ denotes the extra space-like coordinate and $a >0$ is a curvature scalar. Standing waves in the bulk are generated by the oscillatory function:
\begin{equation} \label{u}
u(t,r) = \sin (\omega t) Z(r)~,
\end{equation}
whith
\begin{equation} \label{Z}
Z(r) = |B| e^{-2a|r|} Y_2\left( \frac{\omega}{a} e^{-a|r|} \right)~,
\end{equation}
where $\omega$ denotes the oscillation frequency, $B$ is the integration constant and $Y_2(r)$ is the Bessel function of the second kind of order two.

The oscillatory function (\ref{u}) enters the equations of matter fields via the contravariant components of the metric tensor: $g^{AB}$. We assume that the frequency of standing waves, $\omega$, is much larger than the frequencies associated with the energies of particles on the brane. It allows us to perform the time averaging of the fast oscillating exponents. The non-vanishing time averages are determined by the formula \cite{JHEP}:
\begin{equation} \label{e-u}
\left\langle e^{u} \right\rangle = I_0 \left( Z \right) ~,
\end{equation}
where $I_0(Z)$ is the modified Bessel function of order zero and $Z$ is given by (\ref{Z}). Then the averaged contravariant components of the metric tensor take the form:
\begin{equation} \label{contra}
g^{AB} = \mathrm{diag} \left[e^{-2a|r|}, -I_0(Z)e^{-2a|r|},- I_0(Z)e^{-2a|r|}, -I_0(2Z)e^{-2a|r|}, -1  \right]~.
\end{equation}
Correspondingly, the average of the covariant components of the metric tensor and its determinant are:
\begin{eqnarray}\label{co}
g_{AB} &=& \mathrm{diag}\left[e^{2a|r|}, - \frac{e^{2a|r|}}{I_0(Z)}, -\frac{e^{2a|r|}}{I_0(Z)}, -\frac{e^{2a|r|}}{I_0(2Z)}, -1 \right]~, \nonumber \\
\sqrt G &\equiv& \mathrm{det}(g_{AB}) = \frac{e^{4a|r|}}{I_0(Z)\sqrt{I_0(2Z)}}~.
\end{eqnarray}

The metric (\ref{metric}) contains the increasing warp factor ($a>0$) and it describes the brane located at a node of the standing wave in the bulk (\ref{u}), i.e. the point where the functions $Z (r)$ and $u(t,r)$ vanish. The nodes of the standing wave can be considered as the 4D space-time 'islands', where the matter particles are assumed to be bound. In what follows the replication of fermions families is connected with the existence of such 'islands'.

In our previous papers \cite{JHEP,GMM,GST} we have considered the localization problem in the case of a single node, i.e. when $Y_2(r)$ in (\ref{Z}) has a single zero at the position of the brane, $r = 0$. This was accomplished by the assumption, $\omega/a \approx 3.38$. In this case the 5D Dirac equation has zero mode solutions which correspond to the 4D left fermion stuck at $r=0$ and the right fermion localized in the bulk \cite{GST}. In this paper we consider the model (\ref{metric}) with three nodes and the fine tuned parameters:
\begin{equation}\label{FirstZerosY}
\frac{\omega}{a} \approx 10.02~.
\end{equation}
In this case there appear three additional nodes of the bulk standing wave distributed symmetrically with respect the central node. FIG $1$ shows the shape of the determinant $\sqrt G$ in (\ref{co}) along the extra dimension $r$.

%%%%%%%%%%%%%%%%%%%%%%%%%%%%%%%%%%%%%%%%%%%%%%%%%%%%%%%%%%%%%%%%%%%%%%%%%%%%%
\begin{figure}[ht]
\begin{center}
\includegraphics[width=0.8\textwidth]{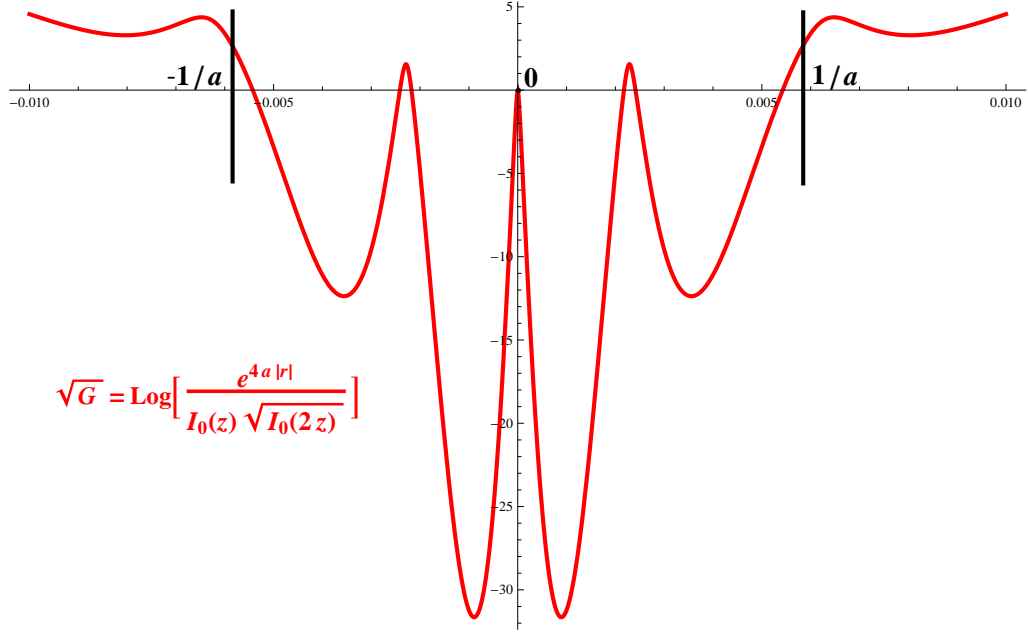}
\caption{Logarithmic profile of the determinant in the bulk}
\end{center}
\end{figure}
%%%%%%%%%%%%%%%%%%%%%%%%%%%%%%%%%%%%%%%%%%%%%%%%%%%%%%%%%%%%%%%%%%%%%%%%%%%%

\noindent
Correspondingly, there appear several fermionic modes which are 'stuck' at different points in the brane of the width $\sim 1/a$, as it is described in \cite{Split}.

The 5D covariant differentiation of the fermion field reads:
\begin{equation} \label{D}
\overset{5}{D}_{M} = \partial_{M}+\frac{1}{4} \Omega_{M}^{\bar{A} \bar{B}} \Gamma_{\bar{A}} \Gamma_{\bar{B}}~,
\end{equation}
where $\Gamma_{\bar{A}}$ are 5D gamma matrices,
\begin{equation}
\Gamma^A = h_{\bar{A}}^A \Gamma^{\bar{A}}~, ~~~
\Gamma^{\bar{A}} = \left( \gamma^0,\gamma^1,\gamma^2,\gamma^3,i \gamma^5 \right)~,
\end{equation}
and
\begin{equation} \label{h}
h_A^{\bar{A}} = \mathrm{diag} \left[ e^{a|r|}, \frac {e^{a|r|}}{\sqrt{I_0(Z)}}, \frac {e^{a|r|}}{\sqrt{I_0(Z)}}, \frac {e^{a|r|}} {\sqrt{I_0(2Z)}}, 1 \right]
\end{equation}
is the {\it f\"unfbein} corresponding to the averaged metric (\ref{contra}).

The spin connection in (\ref{D}) is given by:
\begin{eqnarray} \label{omega}
\Omega_{M}^{\bar{A} \bar{B}} &=& \frac12 \left(h^{N \bar{A}} \eta^{\bar{C} \bar{B}} - h^{N \bar{B}} \eta^{\bar{C} \bar{A}} \right) \partial_M h_{\bar{C} N} - \frac12 \left(h^{N \bar{A}} \eta^{\bar{C} \bar{B}} - h^{N \bar{B}} \eta^{\bar{C} \bar{A}} \right) \partial_N h_{\bar{C} M} + \nonumber\\
&+& \frac12 \eta^{\bar{C} \bar{K}} \left( h^{N \bar{A}} h^{\bar{B} R} -h^{N \bar{B}} h^{\bar{A} R} \right)  h_{\bar{C} M} \partial_R h_{N \bar{K}} ~,
\end{eqnarray}
where $\eta^{\bar{C} \bar{B}}$ is the metric tensor of 5D Minkowski space-time. The non-zero components of (\ref{omega}) for (\ref{h}) are:
\begin{eqnarray}
\Omega_{0}^{\bar{5} \bar{0}} &=& a~ \mathrm{sign}(r)e^{a|r|} ~. \nonumber \\
\Omega_{3}^{\bar{5} \bar{3}} &=& \left[ \mathrm{sign}(r)\frac{a}{\sqrt{I_0(2Z)}} - \frac{I_1(2Z)}{\sqrt{I_0^3(2Z)}} Z' \right] e^{a|r|} ~, \\
\Omega_{1}^{\bar{5} \bar{1}} &=& \Omega_{2}^{\bar{5} \bar{2}} = \left[ \mathrm{sign}(r) \frac{a}{\sqrt{I_0(Z)}} - \frac{I_1(Z)}{2\sqrt{I_0^3(Z)}} Z'\right] e^{a|r|}~, \nonumber
\end{eqnarray}
where primes denote derivatives with respect to $r$.

Let us consider the action of 5D $U(1)$-Higgs model on the standing wave background,
\begin{eqnarray} \label{S}
S &=& \int d^5 x \sqrt{G} \Bigg{[} \frac{i}{2} \overline{\Psi} ~\Gamma^{M} \left( \overset{5}{D}_{M}+ig A_M \right) \Psi - \frac{i}{2}\left( \overset{5}{D}_{M} - ig A_M \right) \overline {\Psi}~ \Gamma^{M} \Psi - \varphi^* \varphi \overline{\Psi} \Psi + \nonumber \\
&+&\left( \partial_M - ig A_M \right)\varphi^* \left( \partial^M + ig A^M \right)\varphi + \frac{\theta^2}{2}\varphi^* \varphi - \frac{\lambda^4}{4} \varphi^{*2} \varphi^2 - \frac{1}{4} F_{M N} F^{M N}\Bigg{]}~.
\end{eqnarray}
We can use the Higgs mechanism to make the 5D fermions massive. After the symmetry braking there still remains a freedom to choose the vector field $A^M$ to be pure gauge, so that it will no longer appear in the action:
\begin{eqnarray} \label{SH}
S &=& \int d^5 x \sqrt{G} \left[ \frac{i}{2} \overline{\Psi} \Gamma^{M} \overset{5}{D}_{M} \Psi - \frac{i}{2} \overset{5}{D}_{M}  \overline{\Psi } \Gamma^{M} \Psi - \right. \nonumber \\
&-& \left. M \overline{\Psi} \Psi + \frac{1}{2} \partial_M \chi \partial^M \chi - \frac{\theta^2}{2} \chi^2 + {\rm Interacting~terms} \right] ,
\end{eqnarray}
where $\chi$ denotes the dynamical 5D Higgs field and the notation: $M=\theta^2/\lambda^4$ was introduced.

Let us use the chiral decomposition of the 5D fermion:
\begin{eqnarray}
\Psi\left( x^A \right) = \psi_L \left( x^\mu \right) \lambda\left( r \right)+ \psi_R \left( x^\mu \right) \rho\left( r \right)~, \nonumber \\
\gamma^5 \psi_R =\psi_R~, ~~~~~~~~~~ \gamma^5 \psi_L =-\psi_L~,
\end{eqnarray}
where $\lambda(r)$ and $\rho(r)$ are the extra dimensional factors of the 4D left and right fermion wavefunctions, $\psi_L \left( x^\mu \right)$ and $\psi_R \left( x^\mu \right)$. Then the fermionic part of (\ref{SH}) takes the form:
\begin{eqnarray}\label{action2}
S_\psi &=& i \int dr \sqrt{G}\frac{\lambda^2}{2} \int d^4 x \left(\overline{\psi}_L \Gamma^\mu D_\mu \psi_L - D_\mu \overline{\psi}_L \Gamma^{\mu} \psi_L \right) + \nonumber \\
&+& i \int dr \sqrt{G}\frac{\rho^2}{2} \int d^4 x \left(\overline{\psi}_R \Gamma^\mu D_\mu \psi_R - D_\mu \overline{\psi}_R \Gamma^{\mu} \psi_R \right) + \\
&+& i \int dr \sqrt{G}\frac{\lambda \rho}{2} \int d^4 x \left(\overline{\psi}_L \Gamma^\mu D_\mu \psi_R + \overline{\psi}_R \Gamma^\mu D_\mu \psi_L- D_\mu \overline{\psi}_R \Gamma^{\mu} \psi_L- D_\mu \overline{\psi}_L \Gamma^{\mu} \psi_R \right) - \nonumber \\
&-& \int dr \sqrt{G}\left( M \rho \lambda -\frac{\rho \lambda'}{2} + \frac{\lambda \rho'}{2} \right) \int d^4 x \left(\overline{\psi}_L \psi_R+\overline{\psi}_R \psi_L \right)~, \nonumber
\end{eqnarray}
where $D_\mu $ denotes the 4D covariant differentiation.

Let us write the 4D Dirac equations for the particles of mass $m$ with the momentum $\overrightarrow{P}=0$ along the brane:
\begin{equation}
P_0 \psi_R = m \psi_L~, ~~~~~ P_0 \psi_L = m \psi_R~.
\end{equation}
Then the 5D Dirac equation yields:
\begin{equation}\label{gant1}
\left\{ \begin{array}{c}
\rho'(r) - m e^{-a|r|} \lambda(r) + \left[ K + M \right] \rho(r) = 0 ~,\\
\lambda'(r) + m e^{-a|r|} \rho(r) + \left[ K - M \right] \lambda(r) = 0~,
\end{array}\right.
\end{equation}
where we use the notation:
\begin{equation}
K = 2a ~ \mathrm{sign}(r) - \frac 12\left[ \frac{I_1(Z)}{I_0(Z)}+\frac{I_1(2Z)}{I_0(2Z)} \right] Z'~.
\end{equation}
Near to the origin $r \approx 0$ the solutions of (\ref{gant1}) are:
\begin{eqnarray} \label{lambda,rho}
\lambda(r) &=& \Bigg{[} \left(C_u \frac{M}{\mu} - C_d \frac{m}{\mu} \right) \sinh(\mu r) + C_u \cosh(\mu r) \Bigg{]}e^{-2ar} ~, \nonumber \\
\rho(r) &=& \Bigg{[} \left(C_u \frac{m}{\mu} - C_d \frac{M}{\mu} \right) \sinh(\mu r) + C_d \cosh(\mu r) \Bigg{]}e^{-2ar} ~, 
\end{eqnarray}
where $C_u$ and $C_d$ are the integration constants and
\begin{equation}\label{mu}
\mu \equiv \sqrt{M^2 - m^2}~.
\end{equation}

To have a localized field on a brane, the 'coupling' constants which appear upon integrating the Lagrangian of this field over $r$ must be non-vanishing and finite. As it was shown in \cite{GST}, the solutions (\ref{lambda,rho}) lead to the localization of 4D fermions on the brane. Indeed, the first two terms in (\ref{action2}) are convergent over $r$ and the third term vanishes, because $\lambda (r) $ is an even and $\rho (r)$ is an odd function of $r$.

%%%%%%%%%%%%%%%%%%%%%%%%%%%%%%%%%%%%%%%%%%%%%%%%%%%%%%%%%%%%%%%%%%%%%%%%%%%%%%
\begin{figure}[ht]
\begin{center}
\includegraphics[width=0.8\textwidth]{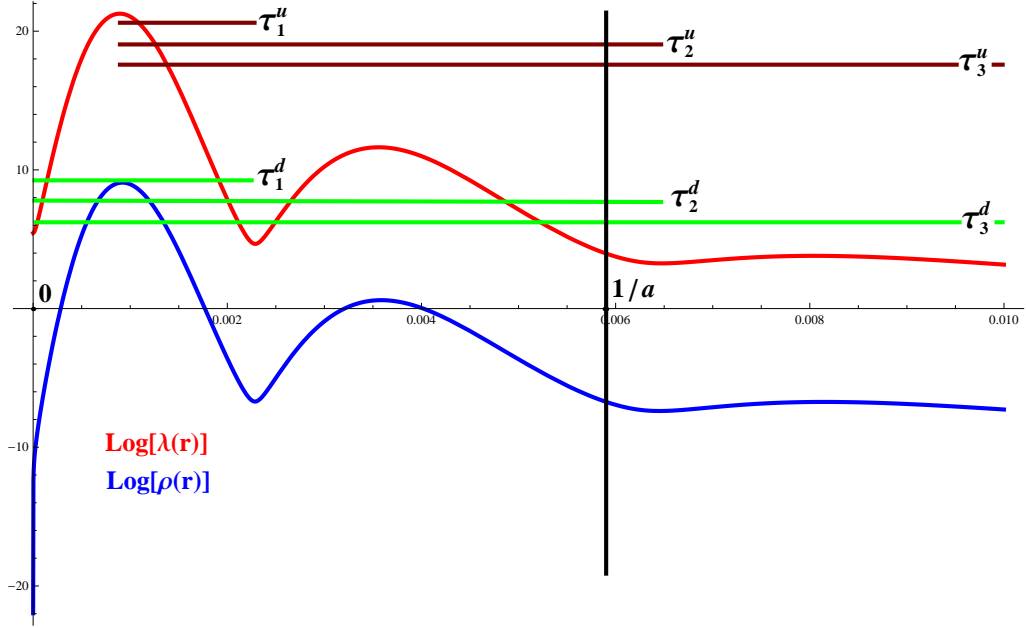}
\caption{Logarithmic profiles of the left and right fermions at the right side from the brane of width $\sim 1/a$}
\end{center}
\end{figure}
%%%%%%%%%%%%%%%%%%%%%%%%%%%%%%%%%%%%%%%%%%%%%%%%%%%%%%%%%%%%%%%%%%%%%%%%%%%%

Now, let us consider the last term of the action (\ref{action2}) which corresponds to the 4D fermion masses:
\begin{eqnarray} \label{m}
m^{fer} &=& \int dr \left( M \rho\lambda + \frac{\lambda \rho'}{2}-\frac{\rho \lambda'}{2} \right) \frac{e^{4a|r|}}{I_0(Z) \sqrt{I_0(2Z)}}= \nonumber \\
&=& \int_0^{\infty} dr \left(\lambda \rho'-\rho \lambda' \right) \frac{e^{4a|r|}}{I_0(Z) \sqrt{I_0(2Z)}}= \\
&=& \left[\left(C_u^2 + C_d^2\right)m - 2 C_uC_dM\right]\int_0^{\infty} \frac{dr}{I_0(Z) \sqrt{I_0(2Z)}} ~.\nonumber
\end{eqnarray}

We connect the fermion families with the existence of several peaks of wave functions which are located at different points in the bulk, and use the fine tuning (\ref{FirstZerosY}) to ensure that $\lambda (r) $ and  $\rho (r)$ acquire three peaks (see FIG $2$). Thus we obtain several dimensional parameters in the model: the distances between the peaks and nodes of the left and right fermion wavefunctions, which can be used to explain the mass spectrum of fermions. The fermion masses are obtained from (\ref{m}), where we impose different limits of integration in order to take into account the overlaps of various left and right fermionic modes.

For example, for three down quarks ($C_u=0$ in (\ref{m})),
\begin{equation}\label{mi}
m_i^d = m_d ~ C_d^2 \int_0^{\tau_i^d} \frac{dr}{I_0(Z) \sqrt{I_0(2Z)}}~,
\end{equation}
the integration limits $\tau_i^d$ ($i=1,2,3$, green lines on FIG $2$) are defined as the distances from the first peak of the left wavefunction, $\lambda (r)$, to the first, second and third nodes of the right wavefunction $\rho(r)$, respectively. From the experiments we know:
\begin{eqnarray} \label{m_d}
m_d &\simeq& 4.5 ~MeV~, \nonumber \\
m_s &\simeq& 100 ~MeV~, \\
m_b &\simeq& 4200 ~MeV~,\nonumber
\end{eqnarray}
and the mysterious large mass ratios:
\begin{equation} \label{frac}
\frac{m_s}{m_d} \simeq 25~, ~~~~~ \frac{m_b}{m_d} \simeq 950~,
\end{equation}
have to be explained. If the parameters of our model are fixed so that
\begin{equation} \label{parameters}
M \simeq a = 170 ~, ~~~ B = 90 ~,
\end{equation}
then from FIG $2$ we see that the upper limits of integration: $\tau_i^d$ in (\ref{mi}),  acquire the values:
\begin{equation}
\tau_1^d = 0.0021~, ~~~ \tau_2^d = 0.0065~, ~~~\tau_3^d = 0.0750~.
\end{equation}
Then from (\ref{mi}) we obtain the masses of down quark families which are consistent with the observed ratios (\ref{frac}).

In the case of up quarks:
\begin{eqnarray}\label{m_u}
m_u &\simeq& 2.5 ~ MeV~, \nonumber \\
m_c &\simeq& 1250 ~ MeV~, \\
m_t &\simeq& 172000 ~ MeV~,\nonumber
\end{eqnarray}
the mass ratios between the generations are:
\begin{equation} \label{ratios}
\frac{m_c}{m_u} \simeq 500~, ~~~~~ \frac{m_t}{m_u} \simeq 70000~.
\end{equation}
Now let us assume $C_d=0$ in  (\ref{m}) and write
\begin{equation}\label{mi2}
m_i^u = m_u ~ C_u^2 \int_{0.0009}^{\tau_i^u} \frac{dr}{I_0(Z) \sqrt{I_0(2Z)}}~,
\end{equation}
where the limits of integration, $\tau_i^u$, are the second peak of $\lambda (r)$ at $r = 0.0009$ and the first, second and third nodes of $\rho (r)$, respectively (see brown lines on FIG $2$), i.e.
\begin{equation}
\tau_1^u = 0.0021~, ~~~ \tau_2^u = 0.0065~, ~~~\tau_3^u = 0.2250~.
\end{equation}
Then, by using the values of the parameters (\ref{parameters}), we are able to obtain from (\ref{mi2}) the observed mass ratios of up quarks in (\ref{ratios}).

As a consequence of our choice of the parameters (\ref{parameters}), the brane width takes the value,
\begin{equation}
\frac 1a = \frac {1}{170} \approx 0.0059~,
\end{equation}
and, as it is clear from FIG $1$, the third node of the standing wave is located outside the brane. From FIG $2$ we see that three peaks of the left fermion wavefunction $\lambda (r)$ are inside the brane, though, the first one, which is located at $r=0$, does not appear on FIG $2$ due to the logarithmic scale. The logarithmic scale on FIG $1$ and FIG $2$ has to be used in order to display the complete structure of the peaks and nodes when the amplitudes of oscillations close to the origin are large. At the same time, all three peaks of the right fermion wavefunction $\rho(r)$ are found in the bulk and the third one is even situated outside of the brane and still overlaps with the left wavefunctions inside the thick brane.

Thus, we have explained the mass ratios of up and down quarks using the dimensionless notations (\ref{parameters}). To obtain the mass spectrum we first need to specify the physical  units. According to (\ref{mu}) the 5D fermionic mass, $M$, which is taken to be of the order of the curvature scale $a$ in (\ref{parameters}), should exceed the 4D mass of any quark.
Hence, in the minimal model we can take
\begin{equation}
M \simeq a \simeq m_t \simeq 172~GeV~.
\end{equation}
Then  the observed mass spectrum of all down and up quarks in (\ref{m_d}) and (\ref{m_u}) can be reproduced if
\begin{equation}
C_d \simeq 860~GeV^{1/2}~, ~~~~~ C_u \simeq 181~GeV^{1/2}
\end{equation}
in (\ref{mi}) and (\ref{mi2}).

To conclude, we considered the 5D standing wave braneworld with three nodes of the standing wave in the bulk. The left and right fermionic wavefunctions on this background have three sharp peaks at different points in the extra dimension. We argue that these peaks can be connected with the fermion families and show that the observed mass spectrum of up and down quark families can be obtained by the proper choice of the free parameters of the model.

%%%%%%%%%%%%%%%%%%%%%%%%%%%%%%%%%%%%%%%%%%%%%%%%%%%%%%%%%%%%%%%%%%%%%%%%%%%%%%%

\section*{Acknowledgments}

This research was supported by the grant of Shota Rustaveli National Science Foundation $\#{\rm DI}/8/6-100/12$.

%%%%%%%%%%%%%%%%%%%%%%%%%%%%%%%%%%%%%%%%%%%%%%%%%%%%%%%%%%%%%%%%%%%%%%%%%%%%%%


\begin{thebibliography}{99}

\bibitem{Family} D.D. Froggatt and H.B. Nielsen,
                Nucl.Phys. {\bf B 147} (1979) 277.

\bibitem{Split} E.A. Mirabelli and M. Schmaltz,
               Phys. Rev. {\bf D 61} (2000) 113011,
               arXiv: hep-ph/9912265;\\
                N. Arkani-Hamed and M. Schmaltz,
               Phys. Rev. {\bf D 61} (2000) 033005,
               arXiv: hep-ph/9903417.

\bibitem{m-mix} Y. Grossman and M. Neubert,
               Phys. Lett. {\bf B 474} (2000) 361,
               arXiv: hep-ph/9912408; \\
                T. Gherghetta and A. Pomarol,
               Nucl. Phys. {\bf B 586} (2000) 141,
               arXiv: hep-ph/0003129; \\
                D.E. Kaplan and T.M. Tait,
               JHEP {\bf 11} (2001) 051; \\
                B.A. Dobrescu and E. Poppitz,
               Phys. Rev. Lett. {\bf 87} (2001) 031801,
               arXiv: hep-ph/0102010; \\
                A. Neronov,
               Phys. Rev. {\bf D 65} (2002) 044004,
               arXiv: gr-qc/0106092; \\
                S.J. Huber,
               Nucl. Phys. {\bf B 666} (2003) 269,
               arXiv: hep-ph/0303183; \\
                M. Gogberashvili, P. Midodashvili and D. Singleton,
               JHEP {\bf 08} (2007) 033,
               arXiv: 0706.0676 [hep-th]; \\
                Z. Guo and B. Ma,
               JHEP {\bf 09} (2009) 091,
               arXiv: 0909.4355 [hep-ph]; \\
                A. J. Buras, C. Grojean, S. Pokorski and R. Ziegler,
               JHEP {\bf 08} (2011) 028,
               arXiv: 1105.3725.

\bibitem{Wave} M. Gogberashvili and D. Singleton,
              Mod. Phys. Lett. {\bf A 25} (2010) 2131,
              arXiv: 0904.2828 [hep-th]; \\
               M. Gogberashvili, A. Herrera-Aguilar and D. Malag\'on-Morej\'on,
              Class. Quant. Grav. {\bf 29} (2012) 025007,
              arXiv: 1012.4534 [hep-th]; \\
               M. Gogberashvili, A. Herrera-Aguilar, D. Malag\'on-Morej\'on, R.R. Mora-Luna and U. Nucamendi,
              Phys. Rev. {\bf D 87} (2013) 084059,
              arXiv: 1201.4569 [hep-th]; \\
               M. Gogberashvili, A. Herrera-Aguilar, D. Malag\'on-Morej\'on and R.R. Mora-Luna,
              Phys. Lett. {\bf B 725} (2013) 208,
              arXiv: 1202.1608 [hep-th].

\bibitem{JHEP} M. Gogberashvili,
              JHEP {\bf 09} (2012) 056,
              arXiv: 1204.2448 [hep-th]; \\
               M. Gogberashvili, P. Midodashvili and L. Midodashvili,
              Int. J. Mod. Phys. {\bf D 21} (2012) 1250081,
              arXiv: 1209.3815 [hep-th].

\bibitem{GMM} M. Gogberashvili, P. Midodashvili and L. Midodashvili,
             Phys. Lett. {\bf B 702} (2011) 276,
             arXiv: 1105.1701 [hep-th]; \\
              M. Gogberashvili, P. Midodashvili and L. Midodashvili,
             Phys. Lett. {\bf B 707} (2012) 169,
             arXiv: 1105.1866 [hep-th].

\bibitem{GST} M. Gogberashvili, O. Sakhelashvili and G. Tukhashvili,
             Mod. Phys. Lett. {\bf A 28} (2013) 1350092,
             arXiv: 1304.6079 [hep-th].

\end{thebibliography}
\end{document}